\documentstyle[preprint,aps]{revtex}
\begin{document}
\preprint{Quantum and Semiclassical Optics 7, 489 (1995)}
\title{Measurement models for time-resolved spectroscopy: a comment}
\author{Holger F. Hofmann\cite{*} and G\H{u}nter Mahler \\ Institut f\H{u}r 
theoretische Physik und Synergetik \\ Pfaffenwaldring 57, 70550 Stuttgart, 
Germany}
\maketitle
\begin {abstract}
We present an exactly solvable model for photon emission, which allows us
to examine the evolution of the photon wave function in space and time.
We apply this model to coherent phenomena in three level sytems with 
special emphasis on the effects of the photon detection process.
\end {abstract}

\section{Introduction}
Describing the open system dynamics of a quantum object, the states of
external fields such as the light field are usually represented by some form of 
statistical approximation: for instance, the master equation approach 
assumes an
unchanging thermal state of the external fields, disregarding all effects which
the system might have on them ("bath approximation").

Approaches such as the Monte Carlo wave function method \cite{1}
or the subensemble density matrix \cite{2}
consider measurements of the photons at the instant of their
emission, thereby arriving at an easy to handle description of the open
system dynamics while avoiding a more detailed description of the light field
itself.

However, in typical experimental situations, the light field is our only source of 
information, while the system dynamics must be deduced indirectly from the 
respective  measurement protocols.
Furthermore, the type of measurement we choose to perform on the photons
may influence the dynamics of the system. Therefore, a closer look at  
the processes which allow us to measure the emitted photons may be helpful.
For this purpose we should start with the complete Schroedinger equation of 
the system and the field. The evolution of this state will then be analyzed 
with respect to a representation adopted to given measurement scenarios.

\section{Basic Model}
In the most simple case, we have a localized two level system,
consisting of the excited state \(|E\rangle\) and the ground state
\(|G\rangle\) with a transition frequency \(\omega_0\), and a one dimensional
field with linear dispersion \(\omega = c|k|\) coupled by a local interaction.

The field can be separated into two distinct branches, one with group velocity
c for $k>0$ and one with group velocity -c for $k<0$. In the following, we will
further simplify the problem by considering only the branch with positive 
group velocity, extending it to negative values of $k$ and $\omega$ as shown
in figure (1). The introduction of negative frequencies can be justified on the
ground that, since the emission of photons with zero or negative energy would
violate energy conservation, such processes will only contribute to the 
dynamics on time scales smaller than $1/\omega_0$. Therefore, the amplitudes
of those spurious field states should be small compared to those near $\omega = 
\omega_0$ \cite{10}.

The relevant states for the emission process are the product state of the 
excited system state and the field vacuum, $|E;vac.\rangle$,
and the product states of the ground state and the 1 photon
field states $|G;k\rangle$.
Using these states as basis, the Hamiltonian of our model is
\begin{eqnarray}
\hat{H} &=& \hbar\omega_0 |E;vac.\rangle\langle E;vac.| \nonumber \\
&+& \int_{-\infty}^{+\infty}dk\,\hbar kc |G;k\rangle\langle G;k| \nonumber \\
&+& \int_{-\infty}^{+\infty}dk\,\hbar  \left[g|G;k\rangle\langle E;vac.|
+ g^*|E;vac.\rangle\langle G;k|\right]
\end{eqnarray} 


This Hamiltonian is a simple version of the Hamiltonian used in 
Wigner-Weisskopf theory. Therefore, we can proceed as in \cite{4}. The time
dependent Schroedinger equation is 
\begin{eqnarray}
\frac{d}{dt} \langle E;vac.|\psi(t)\rangle &=& -i\omega_0 \langle E;vac.|\psi(t)
\rangle \nonumber \\ &-& ig^* \int_{-\infty}^{+\infty}dk\,\langle G;k|\psi(t)
\rangle \\ \frac{d}{dt} \langle G;k|\psi(t)\rangle &=& -i|k|c \langle G;k|
\psi(t)\rangle \nonumber \\ &-& ig \langle E;vac.|\psi(t)\rangle 
\end{eqnarray}
If we choose the initial condition $|\psi(t=0)\rangle=|E;vac\rangle$,
the result of the integrated Schroedinger equation for the one photon part is
\begin{eqnarray}
\langle G;k|\psi(t)\rangle = -ig \int_0^tdt'
e^{-ikc(t-t')}\langle E;vac.|\psi(t')\rangle
\end{eqnarray}
This result can be substituted into the equation for the amplitude of the
excited state. By using $b(t):=e^{i\omega_0t}\langle E;vac.|\psi(t)\rangle$
one obtains
\begin{eqnarray}
\frac{d}{dt}b(t) &=& -|g|^2\int_0^\infty dt'\,b(t')\int_{-\infty}^{+\infty}dk\,
e^{-i(kc-\omega_0)(t-t')}
\end {eqnarray}
The integral over k, again, results in
a delta function, and the evolution becomes local in time:
\begin{eqnarray}
\frac{d}{dt}b(t) = -\frac{\pi |g|^2}{c} b(t) := -\frac{\Gamma}{2} b(t) \\[1cm]
\langle E;vac.|\psi(t)\rangle = e^{(-i\omega_0-\Gamma/2)t}
\end{eqnarray} 
The amplitudes calculated for the 1 photon states of our model are:
\begin{equation}
\langle G;k|\psi (t)\rangle = g e^{-ikct} 
\frac{1 - e^{ -\Gamma t/ 2  } e^{ i (kc - \omega_0) t} }
{ kc - \omega_0 + i \Gamma / 2 }
\end{equation}

We can test our assumption that the amplitudes for $k\le 0$ are much smaller 
than those near $k=\omega_0/c$ by calculating the k-space probability densities:
\begin {equation}
|\langle G;k|\psi (t)\rangle|^2 = |g|^2\frac{1-2\cos(kct-\omega_0 t)
e^{-\Gamma t/2}+e^{-\Gamma t}}{(kc-\omega_0)^2 + \Gamma^2/4} 
\end {equation}
For $t>>1/\Gamma$, this is the familiar Lorentzian distribution. Thus, 
for $\omega_0 >>\Gamma$, the contribution of field states with $\omega < 0$
will be negligible. 
%

On smaller timescales, there are contributions from $k=-\infty$ to $k=+\infty$
due to the energy-time uncertainty. In a model with a more realistic dispersion
relation, the k=0 component have a groupvelocity of 0. Consequently, there will
be reabsorption of such components in the real space formulation, which leads to 
an interaction of the system with itself. The main effect of this interaction 
would be an energy shift in the spectrum, comparabel to the lamb shift in 
quantumelectrodynamics.
However, since our model does not include such processes, there is no such energy
shift and the Lorentzian distribution of the emission centers on $\omega_0$.

%

It is now possible to calculate the real space amplitudes by
Fouriertransforming the k space result: 
\begin{equation}
\langle G;x|\psi (t)\rangle = \left\{ \begin{array}{c}
 -i
\sqrt{\frac{\Gamma}{c}}e^{(\Gamma/2+i\omega_0)(x/c-t)}
 \quad\mbox{for}\quad 0<x<ct \\
0 \qquad \mbox{otherwise}\end{array}\right.
\end{equation}
Figure (2) shows the spatiotemporal distribution of the field state. This 
function can be interpreted as the real space probability density
for photon detection.

\section{Applications}
\subsection{Extension to 3 Dimensions}   
In order to apply this result to realistic conditions in photon spectroscopy,
it is necessary to give 
a more detailed interpretation of the $|k \rangle$ states. In typical cases,
the emission will be into an unrestricted three dimensional light field.
Therefore, an infinite
number of degenerate modes with the same absolute value of k is available.
In the plane wave basis with wavevector ${\bf k}=|{\bf k}|{\bf\hat{k}}$ 
and a linear polarization given
by ${\bf e_p}$, the interaction part of the Hamiltonian is
\begin{eqnarray}
\hat{H}_{int} = & \int d|{\bf k}|d{\bf\hat{k}}\sum_{\bf e_p} 
\hbar g({\bf\hat{k}},{\bf e_p}) |G;{\bf k},{\bf e_p}\rangle 
\langle E;vac.| & \nonumber \\ &
+ \hbar g^*({\bf\hat{k}},{\bf e_p})
|E;vac.\rangle\langle G;{\bf k},{\bf e_p}| & 
\end{eqnarray}
However, instead of using this plane wave basis, we may transform to another
basis in which only a single mode per energy level interacts with the system.  
This new mode can be determined directly from the interaction part of the
Hamiltonian:
\begin{eqnarray}
\hat{H}_{int} = &
\int d|{\bf k}|\hbar\bar{g}|G;k\rangle\langle E;vac.|+\hbar\bar{g}
|E;vac.\rangle\langle G;k| & \\[1cm] \mbox{with} & |G;k\rangle =
\int d{\bf\hat{k}} 
\sum_{\bf e_p} \frac{g({\bf\hat{k}},{\bf e_p})}{\bar{g}}
|G;{\bf k},{\bf e_p}\rangle & \\ \mbox{and} & \bar{g}^2=\int d{\bf\hat{k}}
\sum_{\bf e_p} |g({\bf\hat{k}},{\bf e_p})|^2
\end{eqnarray}
In many situations, it may suffice to inspect the symmetry of the system
and of the field. For example, the spherical symmetry of an atomic system
allows us to classify both the system and the field states according to the 
quantum numbers of the angular momentum, l and m (for details, see \cite{5},
\cite{6}). In the following, we 
restrict the atomic states to $|S\rangle$ (l=0) and $|P\rangle$ (l=1).
A dipole transition emits
only photons of $l=1$ with $m=0,\pm 1$. Since the total angular momentum is
conserved, the system must change its quantum numbers l and m accordingly.  

Since for radii much larger than the wavelength spherical waves approach
plane waves, the previous interpretation of the wave function
in real space still
applies, except in the immediate vicinity of the object. Replacing x by r,
the $|r\rangle$ states now describe a photon at a distance r from the system,
while the angular dependence and the polarization of the photon is given by 
the photon state quantum number m. With the z-axis as quantization axis,
the amplitudes are
\begin{eqnarray}
\langle {\bf k},{\bf e_p}|k';m=0\rangle &=& \frac{1}{4\pi}
\delta(|{\bf k}|-k') {\bf e_p}{\bf e_z} \\ \langle {\bf k},
{\bf e_p}|k';m=\pm 1\rangle &=& \frac{1}{4\pi}\delta(|{\bf k}|-k')
\frac{1}{\sqrt{2}}({\bf e_p e_x}\pm i {\bf e_p e_y})
\end{eqnarray}
The condition for transversality of the light field, 
${\bf e_p}{\bf k}=0$, yields the angular dependence of the emitted intensity
characteristic for dipole radiation. 
\subsection{Three Level System} 
A simple application of this model is the standard quantum beat scenario
(three level system) in which
an atom in a magnetic field is excited from the groundstate $|G\rangle =
|S\rangle$ with a short 
laserpulse polarized in a direction orthogonal to the magnetic field. The atom
is left in a superposition of the $m=+1$ and the $m=-1$ sublevels 
of the excited P state $|E_{\pm}\rangle = |P,m=\pm 1\rangle$. 
Since we can apply our theory to both the $\Delta m = +1$ and the $\Delta m
=-1$ transitions separately and since we are dealing with the linear dynamics 
of the Schroedinger equation, we can immediately write down the complete
evolution of both the field and the system in local representation:
\begin{equation}
|\psi(t=0)\rangle = \frac{1}{\sqrt{2}}(|P,m=+1;vac.\rangle +
|P,m=-1;vac.\rangle)
\end{equation}
\begin{eqnarray}
|\psi(t)\rangle &=& \frac{1}{\sqrt{2}}e^{-\Gamma_+ t/2-i\omega_+ t}
|P,m=+1;vac.\rangle \nonumber\\ &+& \frac{1}{\sqrt2}
e^{-\Gamma_- t/2-i\omega_- t}|P,m=-1;vac.\rangle
\nonumber\\ &-& \frac{i}{\sqrt2}\int_0^{ct}(\sqrt{\frac{\Gamma_+}{c}}
e^{(\Gamma_+/2+i\omega_+)(r/c-t)}
|S;r,m=+1\rangle \nonumber\\&+& \sqrt{\frac{\Gamma_-}{c}}
e^{(\Gamma_-/2+i\omega_-)
(r/c-t)}|S;r,m=-1\rangle)\,dr
\end{eqnarray}
Written in the  m basis of the light field and the m basis of the system,
quantum beats do not appear in the evolution of either subsystem. 
A simple transformation to another basis in atom-,
\begin {equation}
|P_x\rangle = \frac{1}{\sqrt2}(|P,m=+1\rangle + |P,m=-1\rangle)
\end{equation}
\begin{equation}
|P_y\rangle = \frac{i}{\sqrt2}(|P,m=+1\rangle - |P,m=-1\rangle)
\end{equation}
and field-states
\begin{equation}
|r,d_x\rangle = \frac{1}{\sqrt2}(|r,m=+1\rangle + |r,m=-1\rangle)
\end{equation}
\begin{equation}
|r,d_y\rangle = \frac{i}{\sqrt2}(|r,m=+1\rangle - |r,m=-1\rangle)
\end{equation}
shows, however, that beats do occur then, contrary to the statement made in
\cite{3},
that orthogonal dipoles should definitely exclude beats.
The wave function written in this basis for the case of $\Gamma_+=\Gamma_-:=
\Gamma$
now reads:
\begin{eqnarray}
|\psi(t)\rangle &=& e^{-\Gamma t/2-i\bar{\omega}t}(cos(\delta\omega t)
|P_x;vac.\rangle + sin(\delta\omega t)|P_y;vac.\rangle) \nonumber \\
&-& i\sqrt{\frac{\Gamma}{c}}\int_0^{ct} e^{(\Gamma/2+i\bar\omega)(r/c-t)}
(cos(\delta\omega(r/c-t))|S;r,d_x\rangle \nonumber
\\ & + &
sin(\delta\omega(r/c-t))|S;r,d_y\rangle)\,dr
\end{eqnarray}
where $\bar\omega = \frac{\omega_+ + \omega_-}{2}$ and $\delta\omega =
\frac{\omega_+ - \omega_-}{2}$. This representation of the evolution clearly 
shows beats in both the system and the field.
Figure (3) shows the real space probability distribution of this linearly
polarized photon. Of course, 
whether beats are observed or not depends on the type of measurement performed.
\subsection{System-Field Correlations}
Another application of this model is the description of entangled states
between field and system if the decay is from a single excited level
$|E\rangle = |S\rangle$
into two (or possibly more) alternative groundstates $|G_{\pm}\rangle =
|P,m=\pm 1\rangle$.
Although the situation 
may at first appear very similar to the quantum beat scenario, we cannot
single out two paths, since the initial state cannot be separated into two 
components, each belonging to only one path. 
A possibility is to view the decay as a single path, leading from the excited
state to a linear combination of the two groundstates, $ \frac{\Gamma_+}
{\sqrt{\Gamma_+^2+\Gamma_-^2}}|P,m=+1\rangle+\frac{\Gamma_-}{\sqrt{\Gamma_+^2
+\Gamma_-^2}}|P,m=-1\rangle $. However, the two groundstates will generally not
be degenerate. Therefore, the state of the system will start to evolve in
time at
the instant of the emission, t-r/c, causing a dependence of the system state
on the field state $|r\rangle$. The measurement of a photon at r implies that 
a time of r/c has elapsed since emission, so that the state of the total system
would be
\begin{eqnarray}
|G;r\rangle &:=& \frac{1}{\sqrt{\Gamma_+^2+\Gamma_-^2}}
(\Gamma_+ e^{i\omega_+r/c}|P,m=+1;r,m=-1\rangle \nonumber \\&+&
\Gamma_- e^{i\omega_-r/c}|P,m=-1;r,m=+1\rangle)
\end{eqnarray}
Thus we can use the fact that our model treats emissions as instantaneous to
replace the ground state with the temporal evolution of this two level
subsystem.

With $\Gamma = \Gamma_+ + \Gamma_-$ we obtain the result of the complete
evolution as
\begin{eqnarray}
|\psi(t)\rangle &=& e^{-\Gamma t/2-i\omega_0 t}|E;vac.\rangle
- i\sqrt{\frac{\Gamma}{c}}\int_0^{ct}
e^{(\Gamma/2+i\omega_0)(r/c-t)}
|G;r\rangle \,dr \nonumber \\ &=& e^{-(\Gamma_+/2 + \Gamma_-/2 +i\omega_0)t}
|S;vac.\rangle \nonumber \\ &-& i\sqrt{\frac{\Gamma_+ +\Gamma_-}{c}}
\int_0^{ct}e^{(\Gamma_+/2 + \Gamma_-/2+i\omega_0)(r/c-t)}\frac{1}{\sqrt{\Gamma_+
^2+\Gamma_-^2}}  \nonumber \\ & &
(\Gamma_+ e^{i\omega_+(r/c-t)}|P,m=+1;r,m=-1\rangle \nonumber \\[1cm] &+&
\Gamma_- e^{i\omega_-(r/c-t)}|P,m=-1;r,m=+1\rangle)\,dr
\end{eqnarray}
The one photon part of this wave function carries a strong system-field
correlation. In the case $\Gamma_+ = \Gamma_-=\Gamma/2$, any linear combination
of $|P,m=+1\rangle$ and $|P,m=-1\rangle$ states
may be observed, depending on the measurement results in the field.
For example, we may transform to the field basis $|r,d_x\rangle, |r,d_y\rangle$
and the atomic basis $|P_x\rangle, |P_y\rangle$ as in section 3.2.:
\begin{eqnarray}
|\psi(t)\rangle  &=& e^{-(\Gamma/2 +i\omega_0)t}
|S;vac.\rangle \nonumber \\ &-& i\sqrt{\frac{\Gamma}{c}}
\int_0^{ct}e^{(\Gamma/2+i\omega_0+i\bar\omega)(r/c-t)}\frac{1}{\sqrt{2}}
\nonumber \\ & &
(\cos(\delta\omega(r/c-t))(|P_x;r,d_x\rangle + |P_y;r,d_y\rangle)
\nonumber \\[.5 cm] &+&
\sin(\delta\omega(r/c-t))(|P_x;r,d_y\rangle - |P_y;r,d_x\rangle))
\,dr
\end{eqnarray}
Again, we use the notation $\bar\omega = \frac{\omega_+ + \omega_-}{2}$
and $\delta\omega =
\frac{\omega_+ - \omega_-}{2}$.
While the total probability of measuring a photon
with $d_x$ polarization does not oscillate, those photon detections 
correlated with an atomic state of $|P_x\rangle$ and $|P_y\rangle$
both display beats as shown in Figure (4). The beats could therefore be 
seen by measuring the atomic state and separating out the 
photon detections accordingly. This seems to be at variance with the 
expectation that in this type of three level system, beats should be absent 
because a measurement of the  atomic state would reveal the decay channel, 
$\omega_+$ or $\omega_-$\cite{11}.
However, measuring the $|P_x\rangle$ and $|P_y\rangle$ states of the atomic 
system reveals only the phase of the beats, preventing the determination of 
the decay channel. Coincidence measurements of this type can therefore be
considered as quantum eraser measurements as described in \cite{7} and \cite{12}
for the case of 2 spatially separated sources of photonscattering
, since the
measurement of $|P_x\rangle$ and $|P_y\rangle$ effectively erases 
the information to be gained by measuring $|P,m=+1\rangle$ and
$|P,m=-1\rangle$.

The two cases described above and in section 3.2, respectively, may
be combined into a single four level system with
a cascade decay leading from an excited level through two (or more) 
intermediate levels to the ground state. This model could thus be used
to describe the quantum beats observed 
in cascade decays \cite {8}. 
The resulting wave function will 
include a two photon part, so it may be necessary to consider the boson
nature of the photons when transforming to another measurement base.

Since the emission process is local in time and there is no interaction 
between the photons, our present model can be adapted to any other scenario
by simply adding the effects of the possible channels of decay.   
In this way, it should be possible to develop a better understanding of 
the role which the optical measurement apparatus plays in quantum measurements.

\section{Measurement Processes}
For actual measurements the atomic object and the field is embedded into a 
dissipative environment, usually a filter and a detector, which, again, may
be specified by its interaction with the light field. 
The type of measurement
process can be included in the form of a projection acting on 
the field part of the wavefunction. The choice of base states used should 
thus be made with a special experimental setup in mind.
For example, a polarizer-detector setup
can be considered as a projective measurement on $|r;d_x\rangle$ states, where
r is the distance between the system and the detector. With the 3-level system
described by equation (23), this would allow us to observe quantum beats.
A more realistic representation
would consider the limited area covered by the detector, meaning
that actually a state of the form $(\sqrt{\sigma}|r,d_x\rangle + 
\sqrt{1-\sigma}|r,l>1\rangle)$ is measured, where $\sigma$ is the proportion of
photons emitted into the angle covered by the detector. $|r,l>1\rangle$
represents a superposition of multipole states with $l>1$. It ensures,
that the measured state exists only in a limited area by 
interfering with $|r,d_x\rangle$ in such a way, that real space components 
outside the emission angles covered by the detector vanish. 

In our model, the field dynamics in real space rigidly move the field states
away from the
system at the speed of light (see Fig.1).
Therefore, the state of the field represents a 
temporal record of the field dynamics at the system.
It does not matter at what distance r we choose to measure, since a shift in
r is equivalent to a delay time $\Delta t=r/c$.
If we are only interested in time resolved
spectroscopy, we may simulate the photon measurement directly at the system.
This corresponds to the measurement approaches of \cite{1} and \cite{2}.

However, as we have
calculated the whole wavefunction of the field, we can also consider frequency
resolved measurements. Most frequency selective filters are based on an
interference between optical paths of different length. This corresponds to a
linear combination of different $|r\rangle$ states of the field. A Michelson
interferometer would produce a measurement of $\frac{1}{\sqrt{2}}(|r_1\rangle
+ |r_2\rangle)$ in the detector, while a Fabry-Perot interferometer measures
$\frac{1}{\sqrt{1-R^4}}\sum_{n=1}^\infty R^{2n}
|r=nd\rangle$, R being the reflectivity and d the 
distance between the panels.  

In the case of cascades, the consecutive decays will produce a many photon
wavefunction. Nevertheless, we can still apply our theory by using the 
many photon real space base $|r_1,r_2,...\rangle$. Coincidence measurements
such as applied in \cite{8} can now be represented by a measurement base
of $|r_0,r_0+c\tau\rangle$, with $\tau$ as the delay time.

\section{Conclusions}
We have presented an exactly solvable model for the unitary temporal
evolution of
the total system-field state. The assumptions used in the model are equivalent
to the approximations of Wigner-Weisskopf theory. However, our approach 
enables us to provide a physical interpretation of these approximations
in terms of locality and constant group velocity.

Also, it is possible to visualize the evolution
of the wave function in space and time, allowing us to take a closer look
both at time resolved spectroscopy and at the effect of interferometric 
filters. This clearly reveals the special role of the measurement process
in quantum mechanics, which is often concealed
by the application of approximations such as the bath approximation or the
omission of correlations.

The possibility of adapting the model to different situations may provide
an opportunity to study the emission process in complex quantum systems
such as molecules and nanostructures.

\newpage

\begin{figure}

\refstepcounter{figure}\label{fig1}%
{\small {\bf Fig.~\thefigure .} Extension of the linear dispersion 
$\omega = |k|c$ (solid line) to
negative Frequencies (dashed line). The dotted line marks the resonant 
frequency of the system.
}%
\end{figure}

\begin{figure}

\refstepcounter{figure}\label{fig2}%
{\small {\bf Fig.~\thefigure .}
Spatial distribution of the probability of photon detection for 
$t=ln2/\Gamma$ (solid line), $t=2ln2/\Gamma$ (dashed line) and 
$t=3ln2/\Gamma$ (dotted line).
}%
\end{figure}

\begin{figure}

\refstepcounter{figure}\label{fig3}%
{\small {\bf Fig.~\thefigure .} 
Spatial distribution of the probability of photon detection for lineary 
polarized photons in a quantum beat experiment.
}%
\end{figure}

\begin{figure}

\refstepcounter{figure}\label{fig4}%
{\small {\bf Fig.~\thefigure .}
Beats correlated with the $|P_x\rangle$ state (solid line) and the 
$|P_y\rangle$ state (dashed line) of the atomic system. The dotted line 
shows the sum of both probabilitydensities, which displays no beats.         
}%
\end{figure}


\begin{thebibliography}{99}
\bibitem{9} The local interpretation of the photon wavefunction is not 
uncritical, since there is no transversal delta function. However, the 
one dimensional model should represent the propagation of a photon in a 
light wave guide quite well. In three dimensions, we transform only the radial
component to real space, while the angular dependence, which ensures
transversality, remains unchanged.

The local interpretation can further be justified on the ground that
the coupling between system and field is local. Since the measurement
apparatus will interact with the field in a similar fashion, it is 
natural to assume real space locality also for this process.
\end{thebibliography}
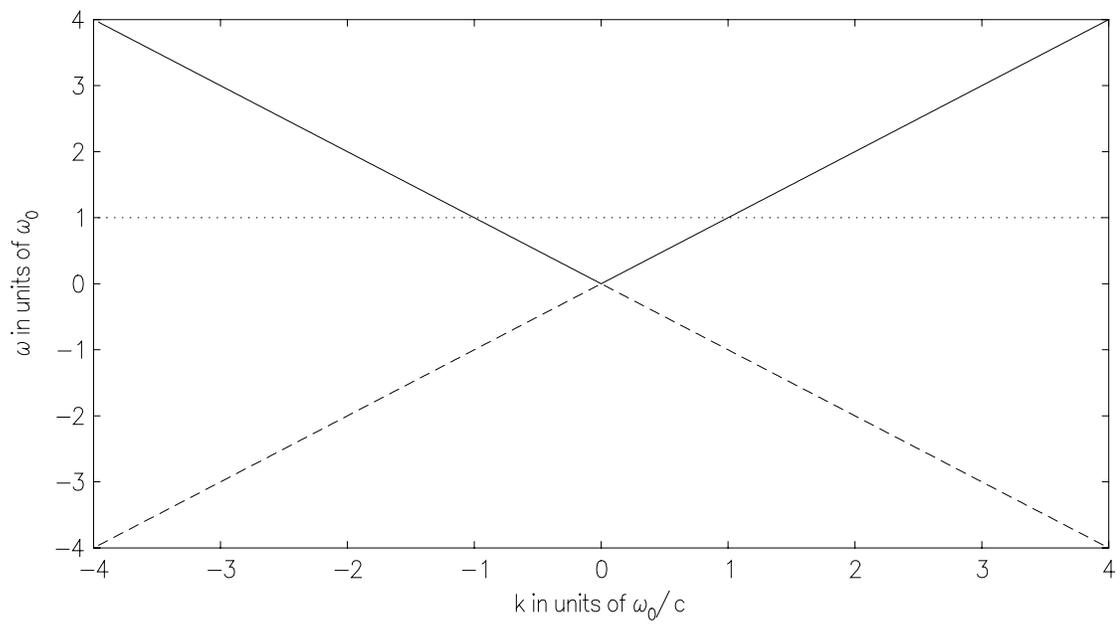
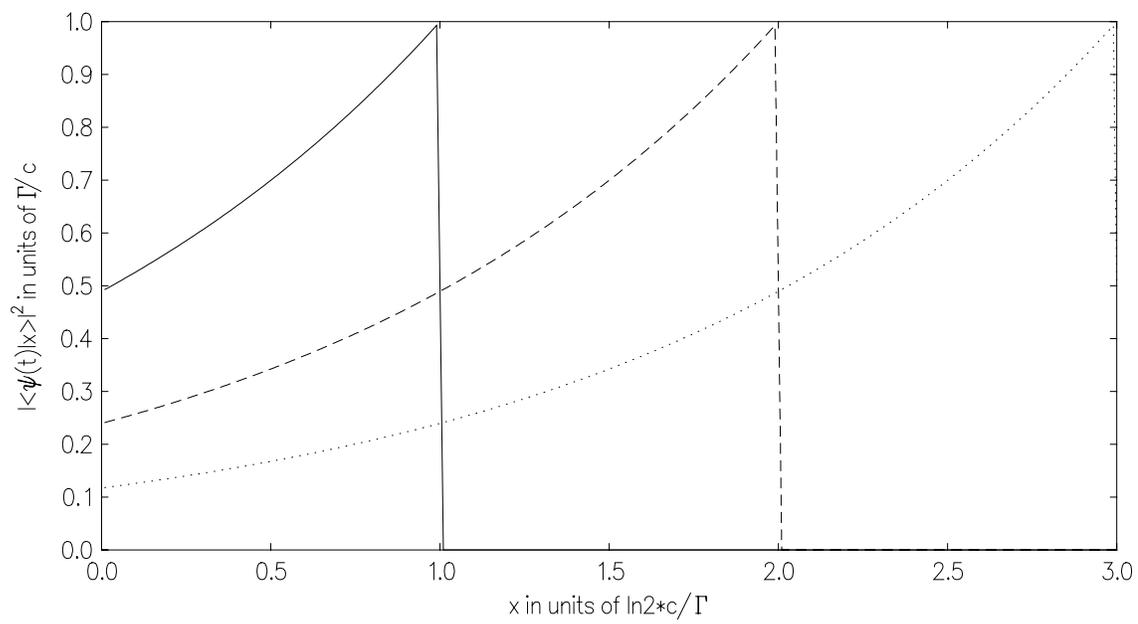
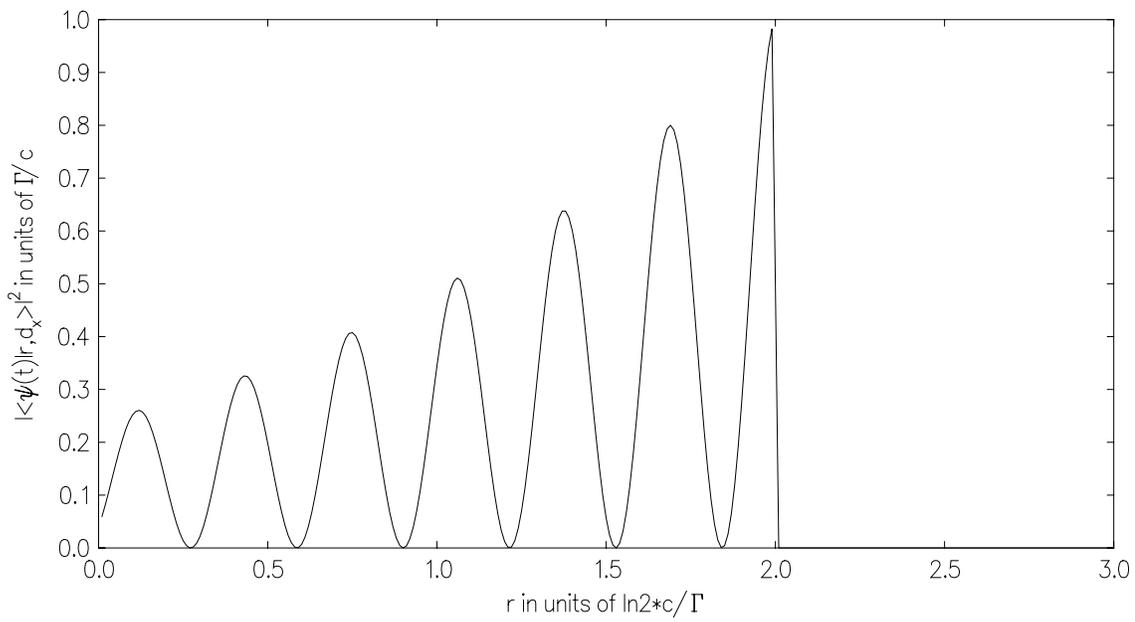
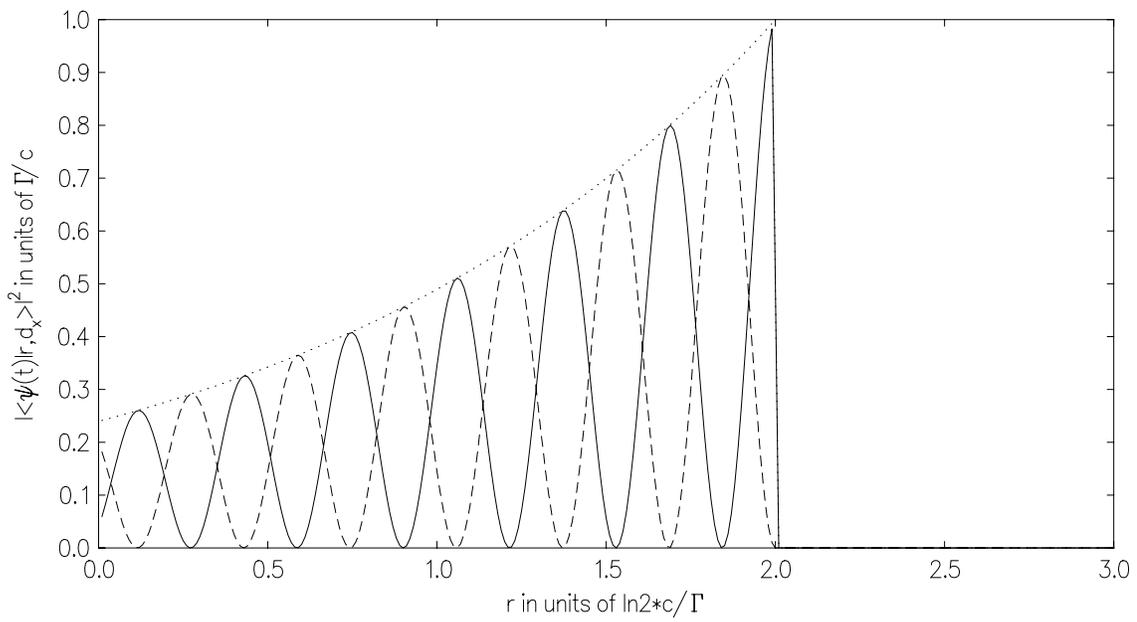
\end{document}